\documentclass[a4paper]{jpconf}
\usepackage{amssymb} 
\usepackage{mathrsfs}
\usepackage{graphicx}
\usepackage{cite}
\usepackage{color}

\def\({\left(}
\def\){\right)}

\newcommand{\R}{\mathbb{R}}

\newcommand{\de}{\textnormal{d}}

\newcommand{\tn}{\textnormal}
\newcommand{\ds}{\displaystyle}

\newcommand{\ie}{\textit{i.e.} }

\newcommand{\citep}[2]{\cite{#1}, p. #2}

\newcommand{\dsfrac}[2]{\ds{\frac{#1}{#2}}}

\newcommand{\mf}[1]{\mathfrak{#1}}
\newcommand{\mc}[1]{\mathcal{#1}}
\newcommand{\mss}[1]{\mathscr{#1}}

\newcommand{\ric}{\tn{Ric}}

\newcommand{\tensors}[3]{\mc T{}^{#1}_{#2}#3}

\newcommand{\metric}[1]{\langle#1\rangle}
\newcommand{\annihg}{\coannih{g}}

\newcommand{\idxannih}[2]{#1{}^{#2}{}}
\newcommand{\idxcoannih}[2]{#1{}_{#2}{}}

\newcommand{\radix}[1]{\idxcoannih{#1}{\circ}}
\newcommand{\annih}[1]{\idxannih{#1}{\bullet}}
\newcommand{\coannih}[1]{\idxcoannih{#1}{\bullet}}

\newcommand{\annihforms}[1]{\annih{\mc A}(#1)}
\newcommand{\fiscal}[1]{\mss F(#1)}
\newcommand{\annihprod}[1]{\annihg(#1)}

\newcommand{\vectmodule}{\mf X}
\newcommand{\fivect}[1]{\vectmodule(#1)}

\newcommand{\fiformk}[2]{\mc A^{#1}(#2)}

\newcommand{\discformsk}[2]{A_d{}^{#1}(#2)}

\newcommand{\srformsk}[2]{\annih{\mss A}{}^{#1}(#2)}

\newcommand{\kosz}{\mc K}
\newcommand{\der}{\nabla}
\newcommand{\dera}[1]{\der_{#1}}
\newcommand{\derb}[2]{\dera{#1}{#2}}

\newcommand{\lder}{\der^{\flat}}
\newcommand{\ldera}[1]{\lder_{#1}}
\newcommand{\lderb}[2]{\ldera{#1}{#2}}
\newcommand{\lderc}[3]{(\lderb{#1}{#2})(#3)}

\newcommand{\cocontr}{{{}_\bullet}}

\newcommand{\image}[3]{\begin{figure*}[ht]
\centering
\includegraphics[width=#2\textwidth]{#1}
\caption{\small{\label{#1}#3}}\end{figure*}}

\newcommand{\abs}[1]{|#1|}

\def\hyph{-\penalty0\hskip0pt\relax}
\newcommand{\rstationary}{radical{\hyph}stationary}
\newcommand{\semiriem}{semi{\hyph}Riemannian}

\newcommand{\semireg}{semi{\hyph}regular}

\newcommand{\quasireg}{quasi{\hyph}regular}

\newcommand{\nondeg}{non{\hyph}degenerate}

\newcommand{\nonsing}{non{\hyph}singular}
\newcommand{\rannih}{radical{\hyph}annihilator}

\newcommand{\FLRW}{FLRW}
\newcommand{\schw}{Schwarzschild}
\newcommand{\rn}{Reissner-Nordstr\"om}
\newcommand{\kn}{Kerr-Newman}

\begin{document}
\title{The geometry of singularities and the black hole information paradox\footnote{Contribution to the 7th Int.  Workshop DICE2014, Sept.  15-19, 2014, Castiglioncello (Italy), see \emph{J.Phys.: Conf. Series}}}

\author{O C Stoica}

\address{Department of Theoretical Physics, National Institute of Physics and Nuclear Engineering -- Horia Hulubei, Bucharest, Romania.}

\ead{cristi.stoica@theory.nipne.ro}

\begin{abstract}
The information loss occurs in an evaporating black hole only if the time evolution ends at the singularity. But as we shall see, the black hole solutions admit analytical extensions beyond the singularities, to globally hyperbolic solutions. The method used is similar to that for the apparent singularity at the event horizon, but at the singularity, the resulting metric is degenerate. When the metric is degenerate, the covariant derivative, the curvature, and the Einstein equation become singular. However, recent advances in the geometry of spacetimes with singular metric show that there are ways to extend analytically the Einstein equation and other field equations beyond such singularities. This means that the information can get out of the singularity. In the case of charged black holes, the obtained solutions have {\nonsing} electromagnetic field. As a bonus, if particles are such black holes, spacetime undergoes dimensional reduction effects like those required by some approaches to perturbative Quantum Gravity.
\end{abstract}

\section{Introduction}


There are two outstanding problems in General Relativity (GR): the problem of singularities, and the quantization of gravity problem.
The theory predicts singularities both inside the black holes, and at the big-bang. It was hoped that they are an artifact of the symmetry of the solutions, but the singularity theorems of Penrose and Hawking show that they are unavoidable \cite{Pen65,Pen69,hawking1970singularities,Haw66i,Haw66ii,Haw67iii}. The occurrence of singularities was regarded from the beginning by many as undesirable for GR, because the field equations don't make sense there, but they are not necessarily that bad. The big-bang singularity is in the past, and is no longer a threat for the fields. The singularities inside the uncharged and non-rotating black holes are spacelike, and they are in the future of any observer that may cross the event horizon, so they are not a problem either. To make harmless the singularities of rotating and charged black holes too, Penrose proposed the {\em cosmic censorship hypothesis} \cite{Pen79,Pen98}, according to which all singularities are either in the past (like the big-bang one) or are always hidden beyond event horizons.

Singularities became more dangerous when Hawking realized that black holes evaporate and the singularity seems to become naked \cite{Haw75,Haw76}. In this case, the information contained in the initial data is partially lost at the singularity, leading to a violation of the unitary evolution. This is the {\em black hole information paradox}.


Hawking evaporation takes place near the event horizon. In fact, particle creation is not limited to the event horizon, but near the horizon the particle-antiparticle pairs may become separated, one of them crossing the horizon and the other one escaping.

Because Hawking evaporation takes place near the horizon, and also because the area of the event horizon is proportional to the entropy of the black hole, it was hypothesized that the event horizon plays a special role, unlike any other light cone. The most investigated approaches to the information loss paradox try to solve the problem at the horizon: this is the case of the {\em black hole complementarity} proposal \cite{susskind1993stretched,susskind2005BlackHoles,susskind2008BlackHoleWar}, and the {\em firewall} proposal \cite{AMPS2013Firewalls,AMPSS2013Firewalls,MP2013GaugeGravityFirewalls}.
However, assigning to the event horizon a special place among other lightlike surfaces in GR is at odds with the {\em principle of equivalence}, especially since we know that black holes evaporate, and the event horizons are only apparent.
In addition, if the information is lost at the singularity, why would the event horizon come to rescue it? It is as if GR has a problem in one place, and we hope to have another problem or perhaps a miracle in another place, so that they cancel one another. If the information is thought to be lost at the singularity, shouldn't we see if we can recover it at the singularity?

We will explore here the possibility that the evolution equations can be extended beyond the singularities, leading to a new possibility, that the information goes out of the singularities.
The main mathematical tool is {\em singular {\semiriem} geometry}, first introduced in \cite{Moi40,Str41,Str42a,Str42b,Str45,Vra42}. It was developed for the case of degenerate metric with constant signature in \cite{Kup87b,Kup96}, but the method was not invariant, requiring the choice of special splits of the tangent bundle. In addition, it could not be extended to metrics with variable signature. The approach I introduced in \cite{Sto11a,Sto11b,Sto11d} overcame these problems. This allowed the extension of the Einstein equation at a large class of singularities \cite{Sto11a,Sto12b}. It could be successfully applied to singularities in GR where the metric is smooth but degenerate, such as the FLRW big-bang model \cite{Sto11h,Sto12a}, leading to a general model of the big-bang which is not necessarily isotropic or homogeneous, but satisfies the Weyl curvature hypothesis \cite{Sto12c}. The more resisting situation was that of black hole singularities, but as I explain below, this case could be worked out too \cite{Sto11e,Sto11f,Sto11g,Sto12e}. In the following, I will show how these results suggest that the information can survive the singularities.

\section{What is the real problem with singularities in General Relativity?}


Let's first identify two different situations when the metric tensor becomes singular. If the metric tensor is smooth, but degenerate ({\ie} it has vanishing determinant), the singularity is called {\em benign}. If the metric tensor has components that become infinite, then the singularity is called {\em malign}.

There are some methods to identify the metric singularities, based on symptoms like geodesic incompleteness, or scalar invariants which become infinite only if the metric is not regular ({\ie} smooth and {\nondeg}). These symptoms are used only to identify singularities, but they don't mean the singularities can't be resolved. The reason why singularities are a problem is not the existence of singular scalar invariants, because there are such invariants which become singular even at points where the metric is regular. Also it is not merely the geodesic incompleteness. These are just symptoms. 

The reason why there are problems at singularities is that the field equations as usually written don't work there. These equations involve the Ricci curvature, and the Levi-Civita connection. The Levi-Civita connection is given in terms of the Christoffel symbols of the second kind, which involve both the components of the metric $g_{ab}$, and of its inverse, $g^{ab}$. For the malign singularities, some of the coefficients $g_{ab}$ are singular, while for the benign ones, they are not, but the inverse $g^{ab}$ is not defined. Hence, for PDE on curved spacetimes the covariant derivatives blow up,
$\Gamma^{\textcolor{red}{c}}{}_{ab} = \ds{\frac 1 2} g^{\textcolor{red}{cs}}(
		\partial_a g_{bs} + \partial_b g_{sa} - \partial_s g_{ab})$,
even in the benign case.
Einstein's equation blows up in addition because the Einstein tensor
\begin{equation}
		G_{ab} = R_{ab} - \dsfrac{1}{2}R g_{ab}
\end{equation}
blows up, being expressed in terms of the curvature, which is defined in terms of the covariant derivative:
$R_{ab} = R^{\textcolor{red}{s}}{}_{asb},\,\, R = g^{\textcolor{red}{pq}}R_{pq}$,
$R^{\textcolor{red}{d}}{}_{abc} = \Gamma^{\textcolor{red}{d}}{}_{ac,b} - \Gamma^{\textcolor{red}{d}}{}_{ab,c} + \Gamma^{\textcolor{red}{d}}{}_{bs}\Gamma^{\textcolor{red}{s}}{}_{ac} - \Gamma^{\textcolor{red}{d}}{}_{cs}\Gamma^{\textcolor{red}{s}}{}_{ab}$.
	
Even if $g_{ab}$ are all finite, these equations are also in terms of $g^{\textcolor{red}{ab}}$, and $g^{ab}\to\infty$ when $\det g\to 0$.

So, if the field equations we normally use are not defined at the singularities, it seems natural to conclude that GR itself predicts its own breakdown, as it was often said \cite{HP70,Haw76,ASH91,HP96,Ash08,Ash09}.

The good news is that the field equations can be written in an apparently equivalent way, but which works also for a large class of singularities. This means that the conclusion that GR predicts its own breakdown was too rushed, and there are still unexplored ways. The secret is to rewrite the equations by replacing the singular fields with {\nonsing} ones, which are equivalent to them as long as the metric is {\nondeg}. Table \ref{table:dictionary} contains a dictionary which allows the replacement of singular fields with {\nonsing} ones for some important classes of singularities.

\begin{table}[htb!]
\begin{center}
   \begin{tabular}{ l | l | l}
     \hline
		 \textbf{Singular} & \textbf{Non-Singular} & \textbf{When g is...} \\ \hline \hline
     $\Gamma^{\textcolor{red}{c}}{}_{ab}$ (2-nd) & $\Gamma_{abc}$ (1-st) & smooth \\ \hline
     \hline
     $R^{\textcolor{red}{d}}{}_{abc}$ & $R_{abcd}$ & {\semireg} \cite{Sto11a} \\ \hline
     $R_{ab}=R^{\textcolor{red}{s}}{}_{asb}$ & $R_{ab}\sqrt{\abs{\det g}}^W,\,W\leq 2$ & {\semireg} \\ \hline
     $R=g^{\textcolor{red}{st}}R_{st}$ & $R\sqrt{\abs{\det g}}^W,\,W\leq 2$ & {\semireg} \\ \hline
     \hline
     $\ric$ & $\ric \circ g$ & {\quasireg} \cite{Sto12b} \\ \hline
     $R$ & $R g \circ g$ & {\quasireg} \\ \hline
     \hline
   \end{tabular}
\end{center}
\caption{A dictionary used to replace singular with {\nonsing} fields.}
\label{table:dictionary}
\end{table}

In Table \ref{table:dictionary}, the symbol $\circ$ denotes the {\em Kulkarni-Nomizu product}, defined as
\begin{equation}
	(h\circ k)_{abcd} := h_{ac}k_{bd} - h_{ad}k_{bc} + h_{bd}k_{ac} - h_{bc}k_{ad}.
\end{equation}

The idea to rewrite the field equations in ways which are equivalent outside the singularity, but are free of infinities at singularities too, is not completely new. In \cite{ASH87,ASH91}, Ashtekar proposed an equivalent Hamiltonian formulation of the Einstein equation, to deal with quantization of gravity, which had some advantages over the ADM formalism \cite{ADM62}. As a side effect, it was believed that rewriting the equations in terms of the ``new variables'' will also resolve the singularities, but this turned out not to be the case for at least one of the variables (see for example \cite{Yon97}). The idea is even older, being proposed in the 1935 article of Einstein and Rosen, who credited W. Mayer for it \cite{ER35}. The idea was to resolve a singularity which was benign, by multiplying Einstein's equations with the determinant of the metric as many times as necessary, and it was only mentioned, and not used explicitly or justified in an invariant way.

\section{Singular {\semiriem} geometry}

In \cite{Sto11a,Sto11b,Sto11d} I extended the mathematical apparatus behind GR, {\semiriem} geometry, to a large class of benign singularities.
We recall here some notions about singular semi-Riemannian manifolds, and some of the main results from \cite{Sto11a}, which will be used in the remainder of the article.

A \textit{singular {\semiriem} manifold} $(M,g)$ is a differentiable manifold $M$ endowed with a symmetric bilinear form $g\in \tensors{0}{2}{M}$ named \textit{metric}. Note that the metric is not required to be {\nondeg}. In particular, if the metric is {\nondeg}, $(M,g)$ is a \textit{{\semiriem} manifold}. If in addition $g$ is positive definite, $(M,g)$ is a  \textit{Riemannian manifold}.

Let $\radix{(T_pM)}=T_pM^\perp$ denote the \textit{radical} of $T_pM$. The metric $g_p$ on $T_pM$ is {\nondeg} if and only if $\radix{(T_pM)}=\{0\}$. The radical of $TM$ is defined as $\radix{T}M=\cup_{p\in M}\radix{(T_pM)}$. We define $\annih{T}M:=\bigcup_{p\in M}\annih{(T_pM)}$, where $\annih{(T_pM)} := (T_pM)^\flat$, and $X_p^\flat(Y_p):=\metric{X_p,Y_p}$, $X_p,Y_p\in T_p M$. The module of sections of $\annih{T}M$ is defined as 
\begin{equation}
	\annihforms{M}:=\{\omega\in\fiformk 1{M}|\omega_p\in\annih{(T_pM)}\tn{ for any }p\in M\}.
\end{equation}
The metric $g$ induces on $\annih{T}M$ a unique {\nondeg} inner product $\annihg$, defined by $\annihprod{\omega,\tau}:=\metric{X,Y}$, where $X^\flat=\omega$, $Y^\flat=\tau$, for two vector fields $X,Y\in\fivect M$.

A tensor $T$ of type $(r,s)$ is called \textit{{\rannih}} in the $l$-th covariant slot if  $T\in \tensors r{l-1}{M}\otimes_M\annih{T}M\otimes_M \tensors 0{s-l}{M}$. The metric $\annihg$ can be used to define uniquely the \textit{covariant contraction} or \textit{covariant trace} between two covariant indices in which a tensor is {\rannih}. If $T\in\tensors r s$ is a tensor field, then we denote the contraction $C_{kl} T$ by
$T(\omega_1,\ldots,\omega_r,v_1,\ldots,\cocontr,\ldots,\cocontr,\ldots,v_s)$.

\textit{The Koszul form} is defined as
$\kosz:\fivect M^3\to\R$,
\begin{equation}
\label{eq_Koszul_form}
	\kosz(X,Y,Z) :=\ds{\frac 1 2} \{ X \metric{Y,Z} + Y \metric{Z,X} - Z \metric{X,Y}
	 - \metric{X,[Y,Z]} + \metric{Y, [Z,X]} + \metric{Z, [X,Y]}\}.
\end{equation}
For {\nondeg} metric, the Koszul form defines the Levi-Civita connection, by raising the $1$-form $\kosz(X,Y,\_)$,
$\metric{\derb X Y,Z} = \kosz(X,Y,Z)$.
Properties of the Koszul form of a singular {\semiriem} manifold are very similar to those of the Levi-Civita connection, and can be found in \cite{Sto11a,Sto11d}. The important difference is that, unlike the Levi-Civita connection, the Koszul form is well defined and finite even if the metric is degenerate.

Let $X,Y\in\fivect M$. The differential $1$-form $\lderb XY \in \fiformk 1{M}$,
$\lderc XYZ := \kosz(X,Y,Z)$
for any $Z\in\fivect{M}$, is called the \textit{lower covariant derivative} of $Y$ in the direction of $X$.
The \textit{lower covariant derivative operator} is defined as
$\lder:\fivect{M} \times \fivect{M} \to \fiformk 1{M}$,
and it associates to each $X,Y\in\fivect{M}$ the differential $1$-form $\ldera XY$.

A singular manifold $(M,g)$ is \textit{{\rstationary}} if, for any $X,Y\in\fivect{M}$,
$\kosz(X,Y,\_)\in\annihforms M$.
Let $X\in\fivect{M}$, $\omega\in\annihforms{M}$, where $(M,g)$ is {\rstationary}. The covariant derivative of $\omega$ in the direction of $X$ is defined as
$\der:\fivect{M} \times \annihforms{M} \to \discformsk 1 M$,
\begin{equation}
	\left(\der_X\omega\right)(Y) := X\left(\omega(Y)\right) - \annihprod{\lderb X Y,\omega},
\end{equation}
where $\discformsk 1 M$ denotes the set of $1$-forms which are smooth on the regions of constant signature.
For a {\rstationary} singular {\semiriem} manifold $(M,g)$, we define:
\begin{equation}
	\srformsk 1 M = \{\omega\in\annihforms M|(\forall X\in\fivect M)\ \der_X\omega\in\annihforms M\}.
\end{equation}

The \textit{Riemann curvature tensor} of a singular {\semiriem} manifold $(M,g)$ is defined as
$R: \fivect M\times \fivect M\times \fivect M\times \fivect M \to \R$,
\begin{equation}
\label{eq_riemann_curvature}
	R(X,Y,Z,T) := (\dera X {\ldera Y}Z - \dera Y {\ldera X}Z - \ldera {[X,Y]}Z)(T)
\end{equation}
for any vector fields $X,Y,Z,T\in\fivect{M}$.

A \textit{{\semireg} {\semiriem} manifold} is defined by the condition
$\ldera X Y \in\srformsk 1 M$,
for any vector fields $X,Y\in\fivect{M}$.
A {\rstationary} {\semiriem} manifold $(M,g)$ is {\semireg} if and only if for any $X,Y,Z,T\in\fivect M$,
$\kosz(X,Y,\cocontr)\kosz(Z,T,\cocontr) \in \fiscal M$.

The Riemann curvature of a {\semireg} {\semiriem} manifold $(M,g)$ is a smooth tensor field $R\in\tensors 0 4 M$. It satisfies for any vector fields $X,Y,Z,T\in\fivect{M}$
\begin{equation}
\label{eq_riemann_curvature_tensor_koszul_formula}
\begin{array}{lll}
	R(X,Y,Z,T)&=& X \kosz(Y,Z,T) - Y \kosz(X,Z,T) - \kosz([X,Y],Z,T)\\
	&& + \kosz(X,Z,\cocontr)\kosz(Y,T,\cocontr) - \kosz(Y,Z,\cocontr)\kosz(X,T,\cocontr).
\end{array}
\end{equation}

The {\em densitized Einstein tensor} $G_{ab}\det g$ is smooth on {\semireg} spacetimes of dimension $4$ even if $G_{ab}$ is singular \cite{Sto11a}. Hence, we can write a {\em densitized Einstein equation}
\begin{equation}
\label{eq_einstein:densitized}
	G_{ab}\sqrt{\abs{\det g}}^W + \Lambda g_{ab}\sqrt{\abs{\det g}}^W = \kappa T_{ab}\sqrt{\abs{\det g}}^W,
\end{equation}
where $T_{ab}$ is the stress-energy tensor, $\kappa:=\dsfrac{8\pi \mc G}{c^4}$, $\mc G$ is Newton's constant and $c$ the speed of light, and $W\in\{0,1,2\}$.
In many cases it is enough to take $W=1$ \cite{Sto11h}, or even $W=0$ \cite{Sto11e}.

\section{Resolving the {\schw} singularity}

The problem of singularities dates back to the very beginnings of GR, when {\schw} wrote his famous solution to Einstein's equation, representing the spacetime metric outside of a spherically symmetric, non-rotating and uncharged body of mass $m$ \cite{Scw16a,Scw16b}. In the coordinates originally found by {\schw}, the metric is
\begin{equation}
\label{eq_schw_schw}
\de s^2 = -\dsfrac{r-2m}{\textcolor{red}{r}}\de t^2 + \dsfrac{r}{\textcolor{red}{r-2m}}\de r^2 + r^2\de\sigma^2,
\end{equation}
where $\de\sigma^2 = \de\theta^2 + \sin^2\theta \de \phi^2$.

{\schw} was dissatisfied with this solution, because the metric appears to be singular at $r=0$ and $r=2m$.
He therefore replaced the coordinate $r$ with $R=r-2m$, and imposed the condition $R>0$. However, the singularity $r=2m$ turned out to be not a singularity of the metric tensor, but of the coordinates. In suitable coordinates, the metric tensor becomes regular at $r=2m$, as shown by Eddington \cite{eddington1924comparison} and Finkelstein \cite{finkelstein1958past}. So the event horizon $r=2m$ should also be included in the solution, together with the interior region of the black hole, $r<2m$.

One major remark about the solution proposed by Eddington and Finkelstein is the understanding that, by moving to another atlas, a metric that looks singular may be made regular. This is not at odds with the principle of general relativity, which states that the physical laws are independent on the coordinates, nor it is at odds with known simple differential geometry observations that the metric remains regular after a non-singular coordinate change, because the coordinate transformations involved here are singular.
In the case of the {\schw} metric, because of the constrained imposed by {\schw} to the solution, the coordinates are singular at $r=0$ and $r=2m$, and we have to move to {\nonsing} ones, which are the correct ones. This is what Eddington and Finkelstein did for the event horizon.

This raises the question whether we can apply the same idea of using a singular coordinate transformation to remove the $r=0$ singularity. This is not the case, because the Kretschmann scalar $R_{abcd}R^{abcd}$ is singular at $r=0$, and no coordinate transformation, not even a singular one can change it, hence either the Riemann tensor $R_{abcd}$ or $R^{abcd}$ is singular in any coordinates, and so is the metric.

However, it is enough to find coordinates in which the singularity is made benign, and not necessarily completely removed.
In \cite{Sto11e} I showed that changing the coordinates to
\begin{equation}
\begin{array}{l}
\bigg\{
\begin{array}{ll}
r &= \tau^2 \\
t &= \xi\tau^4 \\
\end{array}
\\
\end{array}
\end{equation}
leads to the following expression for the four-metric:
\begin{equation}
\label{eq_schw_analytic_tau_xi}
\de s^2 = -\dsfrac{4\tau^4}{2m-\tau^2}\de \tau^2 + (2m-\tau^2)\tau^{4}\(4\xi\de\tau + \tau\de\xi\)^2 + \tau^4\de\sigma^2,
\end{equation}
which is {\em analytic} and {\em \semireg}  at $r=0$.
The metric extends analytically beyond the singularity $r=0$. In the case of evaporating black holes, this leads to the Penrose diagram in Fig. \ref{evaporating-bh-s} B. In the new coordinates, the metric becomes not only analytic, but also {\semireg}. This means that the Einstein equation can be rewritten in a form which holds at the singularity $r=0$ too. 
This suggests that information is not necessarily destroyed by singularities.

\image{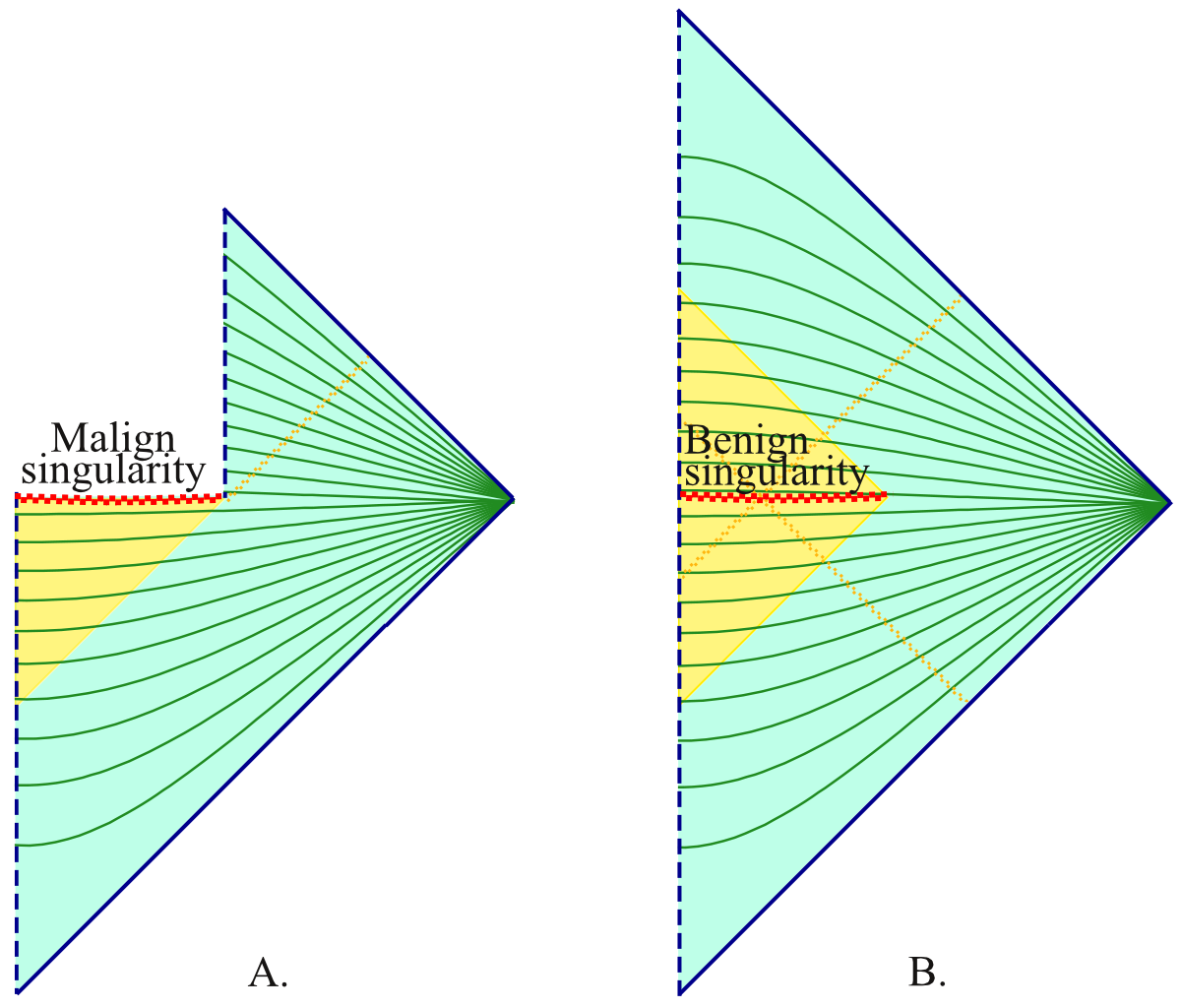}{0.55}{
\textbf{A.} Standard evaporating black hole, whose singularity destroys the information.
\textbf{B.} Evaporating black hole extended through the singularity preserves information.
}

\section{Discussion}

We have seen that a large class of benign singularities (which are due only to the degeneracy of the metric) have reasonable geometric properties, and the equations can be rewritten without infinities. This includes some very general big-bang singularities \cite{Sto11h,Sto12a,Sto12c}. Also, we have seen that the {\schw} black hole singularity, although malign, can be made benign by changing the atlas \cite{Sto11e}.
Similarly, in \cite{Sto11f,Sto11g} I found coordinates that make the metric analytic for {\rn} and {\kn} black holes, and in \cite{Sto12e} that these solutions can be used to obtain globally hyperbolic solutions of stationary and evaporating, charged and neutral, rotating or non-rotating black holes.

Moreover, in the coordinates found for the charged black holes in \cite{Sto11f,Sto11g}, the electromagnetic field and potential are smooth even at the singularities. But these solutions have spherical or cylindrical symmetry. What happens if we assume such a solution as background, and consider other fields that cross the singularity? For example, the Maxwell and the Yang-Mills equations can be written in many equivalent ways, by using the covariant derivative, or using the Hodge dual, or using the codifferential operator $\delta$. All these need the metric to be {\nondeg}, but in \cite{Sto14b} I found conditions which allow us to rewrite these equations without infinities even if the metric becomes degenerate. But the question of what generic initial conditions lead to solutions satisfying such conditions remains open.

Another question concerns the physical interpretation of the formulations mentioned in this article. More precisely, what motivates the preference for the quantities in the second column in Table \ref{table:dictionary}, over those in the first column? As long as the metric is {\nondeg}, these quantities are equivalent. For example, we move from $\Gamma_{abc}$ to  $\Gamma^{c}{}_{ab}$, or from $R_{abcd}$ to  $R^{d}{}_{abc}$, by contracting with $g^{ab}$. So in the {\nondeg} case, the two have the same physical content. But when the metric is degenerate, why would we consider $\Gamma_{abc}$ and $R_{abcd}$ more fundamental than $\Gamma^{c}{}_{ab}$ and $R^{d}{}_{abc}$? Is the fact that they remain finite and smooth at singularities an enough reason? Also, why would we use a version of Einstein's equation in terms of $G_{ab}\sqrt{\abs{\det g}}^W$ and $T_{ab}\sqrt{\abs{\det g}}^W$, $W\in\{0,1,2\}$? If the fact that $\sqrt{\abs{\det g}}^W$ vanishes enough to cancel the divergence of $R_{ab}$ and $R$ is not enough, here are some additional possible reasons, supporting the case $W=1$. First, the Lagrangian density for the Hilbert-Einstein action is $R\sqrt{\abs{\det g}}\de x^0\wedge\de x^1\wedge\de x^2\wedge\de x^3$, and not just $R$. It is a differential form, and not a scalar, and if we insist to consider only the scalar part may lead to infinities at singularities, but the differential form may still remain finite. Also, from the action we obtain the stress-energy tensor density, and not the stress-energy tensor $T_{ab}$. A concrete example can be found in \cite{Sto11h}, where it is explained that at the big bang in the {\FLRW} model, although the scalars $\rho$ and $p$, usually named {\em energy density} and {\em pressure density} become infinite at the singularity, the actual densities are $\rho\sqrt{\abs{\det g}}\de x^0\wedge\de x^1\wedge\de x^2\wedge\de x^3$ and $p\sqrt{\abs{\det g}}\de x^0\wedge\de x^1\wedge\de x^2\wedge\de x^3$, and they remain finite and smooth even at the singularity, and so remain $T_{ab}\sqrt{\abs{\det g}}$, $R_{ab}\sqrt{\abs{\det g}}$, and $R\sqrt{\abs{\det g}}$. Quantities like mass, energy, pressure are defined on extended regions, by integrating such densities, which always have to include in their definitions the quantity $\sqrt{\abs{\det g}}$. On differentiable manifolds we don't integrate scalars, we integrate differential forms, which include together with the scalar, also $\sqrt{\abs{\det g}}$, in the volume form, and remain {\nonsing}. Nevertheless, further investigations of the physical interpretation of the quantities proposed here as more fundamental should be made, to deepen our understanding of why some quantities remain finite and others don't, and which of them are more fundamental than the others, if any.

To conclude, the results mentioned in this article and in the ones cited here suggest that singularities not only are less harmful than it was initially thought, but may have even healthy effects to other problems in GR, like Quantum Gravity \cite{Sto12d}.

\section*{References}
\providecommand{\newblock}{}

\end{document}